\documentclass[sigconf]{acmart}

\AtBeginDocument{%
  \providecommand\BibTeX{{%
    \normalfont B\kern-0.5em{\scshape i\kern-0.25em b}\kern-0.8em\TeX}}}

\setcopyright{acmcopyright}
\copyrightyear{2023}
\acmYear{2023}
\acmDOI{XXXXXXX.XXXXXXX}

\acmConference[CIKM '23]{The 32nd ACM International Conference on Information and Knowledge Management}
{October 21-25, 2023}{University of Birmingham and Eastside Rooms, UK}
\begin{document}


\title{Entire Space Cascade Delayed Feedback Modeling for Effective Conversion Rate Prediction}

\author{Yunfeng Zhao}
\affiliation{
  \institution{School of Software, Shandong University}
  \city{Jinan}
  \country{China}
}
\email{yunfengzhao@mail.sdu.edu.cn}

\author{Xu Yan}
\affiliation{
  \institution{Alibaba Group}
  \city{Hangzhou}
  \country{China}
}
\email{wuyong.yx@taobao.com}

\author{Xiaoqiang Gui}
\affiliation{
  \institution{School of Software, Shandong University}
  \city{Jinan}
  \country{China}
}
\email{x.q.gui@mail.sdu.edu.cn}

\author{Shuguang Han}
\author{Xiang-Rong Sheng}
\affiliation{
  \institution{Alibaba Group}
  \city{Hangzhou}
  \country{China}
}
\email{shuguang.sh@taobao.com}
\email{xiangrong.sxr@taobao.com}

\author{Guoxian Yu}
\affiliation{
  \institution{School of Software, Shandong University}
  \city{Jinan}
  \country{China}
}
\email{gxyu@sdu.edu.cn}

\author{Jufeng Chen}
\author{Zhao Xu}
\author{Bo Zheng}
\affiliation{
  \institution{Alibaba Group}
  \city{Hangzhou}
  \country{China}
}
\email{jufeng.cjf@taobao.com}
\email{changgong.xz@taobao.com}
\email{bozheng@taobao.com}

\renewcommand{\shortauthors}{Zhao and Yan, et al.}

\begin{abstract}
Conversion rate (CVR) prediction is an essential task for large-scale e-commerce platforms. However, refund behaviors frequently occur after conversion in online shopping systems, which drives us to pay attention to effective conversion for building healthier shopping services. This paper defines the probability of item purchasing without any subsequent refund as an effective conversion rate (ECVR). A simple paradigm for ECVR prediction is to decompose it into two sub-tasks: CVR prediction and post-conversion refund rate (RFR) prediction. 
However, RFR prediction suffers from data sparsity (DS) and sample selection bias (SSB) issues, as the refund behaviors are only available after user purchase. Furthermore, there is delayed feedback in both conversion and refund events and they are sequentially dependent, named cascade delayed feedback (CDF), which significantly harms data freshness for model training. Previous studies mainly focus on tackling DS and SSB or delayed feedback for a single event.
To jointly tackle these issues in ECVR prediction, we propose an Entire space CAscade Delayed feedback modeling (ECAD) method. Specifically, ECAD deals with DS and SSB by constructing two tasks including CVR prediction and conversion \& refund rate (CVRFR) prediction using the entire space modeling framework. In addition, it carefully schedules auxiliary tasks to leverage both conversion and refund time within data to alleviate CDF.
Experimental results on the offline industrial dataset and online A/B testing demonstrate the effectiveness of ECAD. In addition, ECAD has been deployed in one of  the recommender systems in Alibaba, contributing to a significant improvement of ECVR.

\end{abstract}






\maketitle

\vspace{-0.5em}
\section{Introduction}
\label{sec: intro}
To facilitate an effective matching of user interest with millions of online products, e-commerce platforms usually spend a significant amount of effort on developing deep machine learning models such as the click-through rate (CTR) prediction model~\cite{zhou2018deep,sheng2023joint} and the conversion rate (CVR) prediction model~\cite{yang2021capturing,chan2023capturing}. Given the importance of conversions in online shopping services, we mainly focused on the realm of CVR prediction in this study. Researchers and practitioners have also proposed a variety of approaches for obtaining an accurate estimation of CVR~\cite{wen2020entire,wen2021hierarchically,ma2018entire,xu2022ukd,gu2021real,yang2021capturing}. 

To tackle the problem of data sparsity (DS), Wen et al.~\cite{wen2020entire,wen2021hierarchically} utilized a series of post-click behaviors for supplementing the rare conversion event, which resulted in an improved performance on CVR prediction. Others dealt with the selection bias problem (SSB) in CVR prediction and proposed the entire space estimation method by jointly learning with multiple behavior prediction tasks~\cite{ma2018entire,xu2022ukd}. Moreover, unlike click behaviors, feedback labels for the conversion event cannot be collected within a short time as conversion may happen in days or even weeks afterward. This introduces a dilemma – to employ recent data with incomplete labels or to wait for more accurate labels. By assuming a stable delayed conversion pattern (e.g., the delay time for the final conversion remains steady), a delayed feedback model resolves this problem by utilizing the relationship between the ultimate conversion and the conversion time to correct the incomplete real-time labels~\cite{gu2021real,yang2021capturing}. 
\begin{figure*}[h!tbp]
    \centering
    \vspace{-1em}
    \includegraphics[width=14cm]{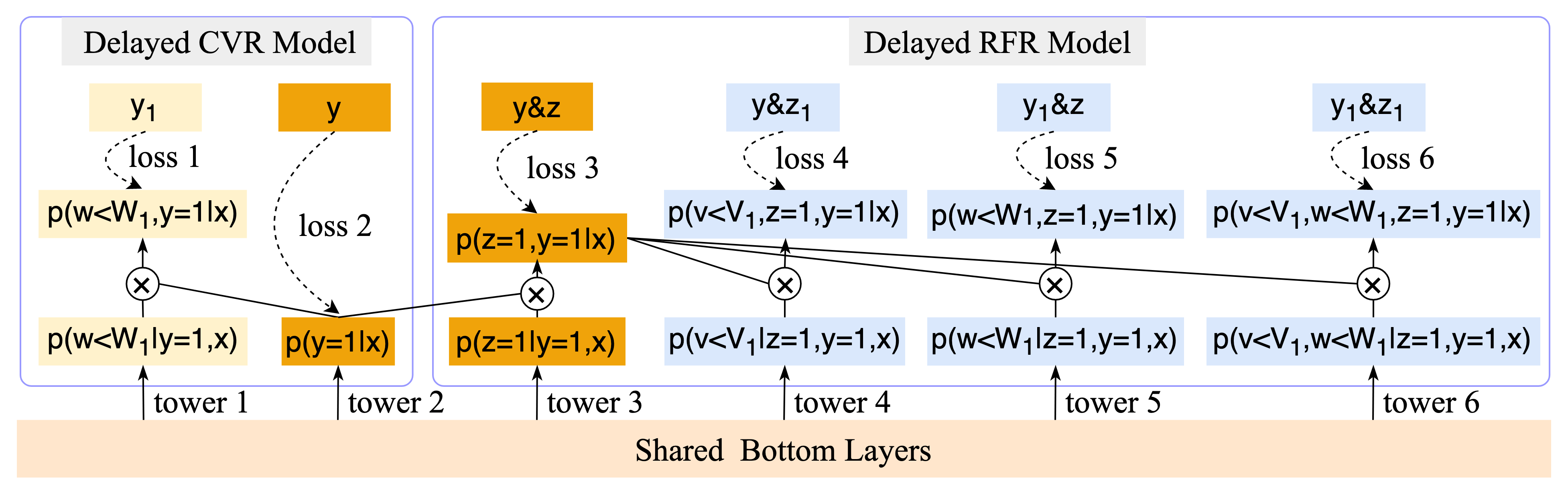}
    \vspace{-1.5em}
    \caption{The overall  schematic framework of ECAD, which consists of three parts: shared bottom layers, a delayed CVR model, and a delayed RFR model. The shared bottom layers take the sample features (i.e. discrete IDs) as input and output a fixed-length vector,  and then feed this vector into the towers of delayed CVR and delayed RFR models to predict the corresponding probabilities. $V_1$ and $W_1$ denote the pre-defined windows for conversion and refund, respectively.}
    \label{fig: ECAD}
    \vspace{-1em}
\end{figure*}

In real-world online shopping systems, we frequently observe refund requests as the purchased products may not truly fulfill users’ real interests. To build a healthy e-commerce platform, we need a better understanding of effective conversions. Considering that user behaviors follow a sequential pattern of $click  \to conversion \to refund$, we define an effective conversion rate (ECVR) prediction problem: to forecast the probability that a product will eventually be purchased without a refund after the user clicks this product. However, both conversion and refund events suffer delay feedback, i.e. the conversion and refund events usually occur over a long period after the click and conversion events happened, respectively.

A canonical manner to obtain the ECVR label is to set two label attribution windows to observe conversion and refund events separately, i.e. a conversion (refund) is attributed to positive if it happens within the corresponding attribution windows since click (conversion) occurs. Then, ECVR prediction can be achieved by directly training a prediction model (e.g. Embedding\&MLP~\cite{zhou2018deep,zhou2019deep} or Wide\&Deep~\cite{cheng2016wide,zhu2021open}) using those samples. However, this strategy significantly compromises data freshness since the corresponding ECVR labels are determined by two attribution windows for observing conversion and refund events. Therefore, the latest data cannot be readily filled into the model, which cannot catch up with the dynamic change of data distribution~\cite{ye2022future,xu2022ukd}. Furthermore, such a naive strategy treats both non-conversion samples and conversion-while-refunded samples with equal importance, ignoring that they possess distinct information for ECVR prediction.

To tackle the above issues, we can simply divide ECVR prediction into two sub-tasks: CVR prediction and post-conversion refund rate (RFR) prediction. Accordingly, the probability of ECVR $p_{ecvr}$ equals to $p_{cvr}*(1-p_{rfr})$, where $p_{cvr}$ and $p_{rfr}$ represent the probabilities of CVR and RFR, respectively. In this way, the attribution window for each sub-task will be shorter compared to that of the ECVR task, alleviating the delayed feedback problem for ECVR prediction to some extent. Despite that, there still exist several major challenges if we simply model CVR prediction on top of the clicked samples, and RFR prediction with the conversion samples. 

Concretely, the first few challenges are \textbf{DS}~\cite{lee2012estimating} and \textbf{SSB}~\cite{zadrozny2004learning}. DS means that the RFR prediction model is built on top of the conversion samples, which are significantly less than the clicked samples (for training the CVR prediction model). As such, the limited supervised information makes it hard for the RFR predictor to learn sufficient knowledge for accurate prediction. SSB refers to the inconsistency of data distribution between the training and the testing space, i.e. the RFR predictor is trained using conversion samples but is deployed to make inferences on the entire space with all clicked samples, which hurts the model generalization ability. 

A common solution for DS and SSB is to employ the entire space modeling approach~\cite{ma2018entire,wang2022escm2}. In our case, we can utilize the clicked samples to jointly train RFR and CVR prediction models, which alleviates the DS and SSB problems for RFR (as mentioned, it is often trained only with the conversion samples). In addition, there is another big challenge, and we name it the \textbf{Cascade Delayed Feedback (CDF)} problem -- both of them have delayed feedback issues, and RFR is further sequentially dependent on the CVR task. Although several approaches \cite{ktena2019addressing,yasui2020feedback,wang2020delayed,yang2021capturing,gu2021real,hou2021conversion,huangfu2022multi,yang2022generalized, chen2022asymptotically,yasui2022learning} have been proposed to tackle delayed feedback in CVR prediction, they only focus on one delay event and thus incapable to address CDF in the entire space modeling framework.

To this end, we introduce an Entire space CAscade Delayed feedback modeling (ECAD) framework for ECVR prediction from the perspective of sequentially modeling CVR and RFR prediction in a sensible way. Specifically, we adopt an entire space modeling framework and construct two tasks: CVR and conversion \& refund rate (CVRFR) prediction to help induce the RFR prediction model. In addition, we carefully design auxiliary tasks for CVR and CVRFR prediction using multi-task learning~\cite{zhang2020large,zhang2021survey} to handle the challenging CDF problem. Fig.~\ref{fig: ECAD} presents  the overall schematic of ECAD and the main contribution of our work are summarized as follows:

\begin{itemize}
\item To the best of our knowledge, this work makes the first step toward modeling ECVR prediction. We carefully divide this prediction into two sequentially dependent sub-tasks, i.e. CVR and RFR prediction, and point out a novel, thorny but untouched challenge of these two tasks, i.e. CDF problem.

\item We propose an Entire space CAscade Delayed feedback modeling (ECAD) approach for learning the ECVR prediction model that adopts an entire space modeling framework to address the SSB and DS problems. Besides, to tackle the CDF problem, ECAD carefully designs auxiliary tasks to leverage the conversion and refund time contained in delayed feedback data to advance the prediction performance.

\item  Experimental results on the offline industrial dataset and online A/B testing clearly demonstrate the effectiveness of ECAD.  Moreover, ECAD has been deployed in one of the production recommender systems in Alibaba, yielding a significant improvement of 5.21\% ECVR.
\end{itemize}

\vspace{-0.25em}
\section{Related Work}
Our work is closely associated with entire space modeling and delayed feedback modeling, and we discuss the previous research in the following two aspects.

\textbf{Entire Space Modeling}. Data sparsity (DS) and sample selection bias (SSB) are two critical problems in recommender systems. In the last decade, many attempts have been made to solve these two challenges. Taking the CVR prediction task in recommender systems as an example, the user actions generally follow a sequential pattern of $impression \to click \to conversion$. The CVR predictor that outputs the post-click conversion rate is usually trained using the clicked samples but makes inferences on the entire space with all impression samples, resulting in SSB. In addition, the clicked samples for training the CVR predictor are scarce compared to impression samples for learning the click-through rate (CTR) one~\cite{richardson2007predicting,zhou2018deep}, bringing DS. 
Early works only focused on a certain challenge, i.e. DS~\cite{weiss2004mining,lee2012estimating,su2021attention} or SSB~\cite{pan2008one,zhang2016bid}. To combat these two challenges simultaneously, 
\citet{ma2018entire} proposed ESMM which utilizes two auxiliary tasks, i.e. CTR and click-through \& conversion rate (CTCVR), to train the CVR model indirectly over the entire space of all impression samples and thus align to sample space of testing. Hence, the DS and SSB are alleviated simultaneously. The superiority of ESMM is further demonstrated in $\text{ESM}^2$~\cite{wen2020entire}, AITM~\cite{xi2021modeling}, $\text{ESCM}^2$~\cite{wang2022escm2}, HEROES~\cite{jin2022multi} and  DCMT~\cite{zhu2023dcmt}. In this paper, we refer to the idea of ESMM to design two tasks (i.e., CVR and CVRFR prediction) for addressing the DS and SSB when learning the RFR prediction model, and then perform the ECVR prediction by the learnt CVR and RFR predictor. Note that we focus on the sequential pattern of actions $click  \to conversion \to refund$, thus the entire space consists of all clicked samples in our task. 
However, the above-mentioned entire space modeling methods ignore the delayed feedback in sequential actions, which significantly harms the data freshness. 

\textbf{Delayed Feedback Modeling}. There usually is a severe delayed feedback in CVR and RFR prediction, due to the fact that the conversion/refund action may occur hours or days later after the click/conversion event, which hurts the data freshness for training the predictor.
A series of works ~\cite{ktena2019addressing,yasui2020feedback,yang2021capturing,gu2021real,huangfu2022multi,yang2022generalized, chen2022asymptotically,yasui2022learning} have been proposed to redesign the data pipeline and loss function to address the delayed feedback problem for online learning scenario with streaming training.
In rough, the fake negatives are first labeled as negative ones to train the model and then duplicated as positive ones to train the model once the conversion happens within the attribution window.
Furthermore, ~\citet{gu2021real} proposed the Defer, which develops an effective offline training method that incorporates multi-task learning to harness the information inherent in conversion time for CVR prediction. However, these works only consider the delayed feedback problem in one event, which is distinct from our task, i.e. both conversion and refund events suffer the issue of delayed feedback, causing the Cascade Delayed Feedback (CDF) problem. 
Here, we adopt multi-task learning and carefully design auxiliary tasks to leverage both conversion and refund time contained in data for handling CDF in ECVR prediction. Note that we focus on how to deal with the CDF problem, but not on proposing an effective delayed feedback model for a single event.

\section{Preliminary}
In this section, we first formulate the problem for ECVR prediction and then introduce how previous offline training methods address the delay feedback issue in detail.
\subsection{Problem Formulation}
\label{sec: Problem Formulation}
Given a dataset $\mathcal{D}= \{(\mathbf{x}, y, z, \hat{y})\}^N$, $(\mathbf{x}, y, z, \hat{y})$ marks a sample and $N$ is the number of clicked samples, where $\mathbf{x}$ denotes the high-dimensional feature vector consisting of multi-fields (e.g. user and item field), $y$ and $z$ are binary labels with $y=1$ or $z=1$ meaning the conversion or refund event occurs respectively, $\hat{y}$ is the binary effective conversion label with $\hat{y} = 1$ indicating the sample is converted and no refunded (i.e. $y=1$ and $z=0$). Note that there is a sequential dependence between conversion and refund labels, i.e. the conversion event always precedes the refund action. Furthermore, both the conversion and refund events suffer from the delayed feedback problem, resulting in the label of many samples cannot be determined even for a long period, i.e. CDF problem. Denote the duration between the click and conversion and between the conversion and refund as $w$ and $v$ respectively, $w (v)=+\infty$ means the sample $\mathbf{x}$ has no conversion (refund) eventually. A commonly-used manner to deal with the delayed feedback problem is setting a time of attribution window to observe the event. Specifically, let $W$ and $V$ be the attribution window of conversion and refund, respectively. If $w<W (v<V)$, the label of conversion (refund) is set to 1.

Our task is to accurately estimate the probability of ECVR $p_{ecvr} = p (\hat{y} = 1|\mathbf{x})$ for the testing sample $\mathbf{x}$. A naive strategy to achieve this problem is learning the prediction model (e.g. Embedding\&MLP~\cite{zhou2018deep} or Wide\&Deep~\cite{cheng2016wide}) with input $(x, \hat{y})$. However, this pattern would arise two concerns: (i) it takes too long for determining the ECVR label $\hat{y}$, i.e. $W+V$, which significantly harms the data freshness for learning the predictor;  and (ii) non-conversion samples and samples that are converted but refunded have different knowledge for modeling ECVR, but they are treated equally in this manner.

To tackle these issues, we transfer the $p_{ecvr} = p (\hat{y} = 1|\mathbf{x}) = p (y=1, z=0|\mathbf{x})$ into two associated probabilities, i.e. probability of  CVR $p_{cvr} = p (y = 1|\mathbf{x})$ and probability of post-conversion RFR $p_{rfr} = p (z=1|y = 1,\mathbf{x})$ in the follows:
\begin{equation}
     \underbrace{p(y=1, z=0|\mathbf{x})}_{p_{ecvr}} = \underbrace{p(y=1 |\mathbf{x})}_{p_{cvr}}  * (1-\underbrace{p( z=1|y=1,\mathbf{x})}_{p_{rfr}}) 
     \label{eq: ecvr}
\end{equation}
In this way, the $p_{ecvr}$ for a given sample can be computed by the estimated  $p_{cvr}$ and $p_{rfr}$.

\subsection{Offline Delayed Feedback Modeling}
CVR and RFR prediction usually suffer from delayed feedback, thus corresponding labels of many samples can not be determined even for a long period. A naive strategy to deal with delayed feedback is to wait for the time of attribution window to ascertain the labels of samples and then utilize them for training the model. However, the data distribution is dynamically changing in recommender systems, e.g. new items or users join the platform. As such, the predictor should also be updated with recent samples near the end of training time to capture the distribution shift in data.

To tackle this problem, \citet{gu2021real} proposed an offline training approach named Defer additionally incorporating the knowledge contained within the recent samples to improve the CVR prediction model. Specifically, Defer employs a multi-task learning framework to advance model generalization utilizing the information included in the conversion time. The model has $n + 1$ towers on top of the shared bottom layers, and an example of $n=1$ is presented in the left part of Fig.~\ref{fig: ECAD}.  One of the towers predicts $p(y=1|\mathbf{x})$, i.e. $p_{cvr}$, while others predict whether the conversion sample will convert within the predefined time windows $W_1, W_2, \cdots, W_n$, i.e. $p(w<W_i|y=1,\mathbf{x})( i:1\to n)$, where $W_i<W (\forall i)$. To update model in space with all samples, $p(w<W_i|y=1,\mathbf{x})$ is further transferred using $p(y=1|\mathbf{x})$ as follows:
\begin{equation}
    p(w<W_i, y=1|\mathbf{x}) = p(w<W_i|y=1,\mathbf{x})*p(y=1|\mathbf{x})
\end{equation}
where $p(w<W_i, y=1|\mathbf{x})$ indicates the predicted probability of the sample $\mathbf{x}$ will convert within the pre-defined time windows $W_i$.
Denoting $y_i$ the label that whether sample $\mathbf{x}$ convert within $W_i$ after click, the loss of Defer is computed as follows:
\begin{equation}
     \mathcal{L} = -\sum_{\mathbf{x}}\sum_i l( p(w<W_i, y=1|\mathbf{x}), y_i)
     -\sum_{\mathbf{x}}l(p(y=1|\mathbf{x}), y )
\end{equation}
where $l(\cdot)$ denotes the binary cross-entropy loss function. Due to recent samples that near the end of the training date may not have all $n+1$ labels, Defer only updates the corresponding parameters according to the observed labels. For a sample, if the predefined $W_{i+1}, \cdots, W_{n}$ windows have not reached, Defer only updates the parameters through losses computed by $p(w<W_1, y=1| \mathbf{x}), \cdots, p(w<W_i, y=1| \mathbf{x})$ and corresponding observed labels $y_1,\cdots,y_i$  while freezing the parameters of other towers.
Note that the tower predicts $p(y=1| \mathbf{x})$ will also be updated through the gradient of losses regarding $p(w<W_1, y=1| \mathbf{x}), \cdots, p(w<W_i, y=1| \mathbf{x})$. For example, assuming the attribution window $W$ as 3 days, and 1 day, 2 days as two extra time windows. For samples clicked on the 1st day from the last, Defer only updates the parameters from $p(w<W_1, y=1| \mathbf{x})$. As to samples clicked before the 2nd day from the last, 
all parameters are updated simultaneously. In addition, the delayed feedback problem in RFR prediction can also be alleviated with Defer using the conversion samples.

\vspace{-0.5em}
\section{The Proposed approach}
In this section, we first give a brief review of ECVR Modeling and Challenges and then detail the proposed Entire space CAscade Delayed feedback modeling (ECAD) method for ECVR prediction.
\subsection{ECVR Modeling and Challenges}
As mentioned in Section~\ref{sec: Problem Formulation}, it is profitable to decompose ECVR prediction into two sub-tasks, i.e. CVR and RFR prediction. An intuitive manner to fulfill this pattern is by constructing two independent models, in which one trained with clicked samples predicts the $p_{cvr} = p(y=1|\mathbf{x})$, while another updated with conversion samples estimates the $p_{rfr} = p(z=1|y=1,\mathbf{x})$. Then those two probabilities are employed for computing the $p_{ecvr} = p(z=1, y=0|\mathbf{x})$ using Eq.~\eqref{eq: ecvr}. However, this strategy for modeling ECVR prediction suffers from several problems (i.e. SSB, DS, and CDF) as mentioned in Section~\ref{sec: intro},  making it sub-optimal.

\subsection{ECAD method}
To address the above issues, we propose an ECAD approach from the perspective of the entire space modeling to construct the CVR and CVRFR prediction tasks for easing the SSB and DS of RFR modeling. In addition, the CDF problem is tackled by incorporating both the knowledge of conversion and refund time contained in data for improving the generalization of models. The overall framework of  ECAD is shown in Fig.~\ref{fig: ECAD}, which consists of three parts: shared bottom layers, a delayed CVR model, and a delayed RFR model. Taking the canonically-used Embedding\&MLP model architecture as an example, the shared bottom layers take the sample features (i.e. discrete IDs) as input and output a fixed-length vector,  and then feed this vector into the towers of delayed CVR and delayed RFR models to predict the corresponding probabilities. The followings elaborate on how we build the ECAD model.

\subsubsection{Prediction  Model}
The target of the prediction model is to estimate $p_{ecvr} = p (y=1, z=0|\mathbf{x})$ by alternatively modeling $p_{cvr} = p(y=1|\mathbf{x})$ and $p_{rfr} = p(z=1|y=1,\mathbf{x})$ using Eq.~\eqref{eq: ecvr}.  However, this manner arises the SSB and DS problems for RFR prediction. To tackle these issues, we adopt the  entire space modeling framework~\cite{ma2018entire} for CVR and RFR prediction. Specifically, by exploiting the sequential dependence of $ click \to conversion \to refund $, we construct CVR and post-click conversion \& refund rate (CVRFR) prediction tasks. Denote $p(z=1,y=1|\mathbf{x})$ as $p_{cvrfr}$ of the sample $\mathbf{x}$, then $p_{rfr}$ can then be formulated as follows:
\vspace{-0.5em}
\begin{equation}
    p(z=1|y=1,\mathbf{x}) = \frac{p(z=1,y=1|\mathbf{x})}{p(y=1|\mathbf{x})}
    \label{eq: rfr}
\end{equation}
where $p(z=1,y=1|\mathbf{x})$ and $p(y=1|\mathbf{x})$ are modeled on the entire space with click samples, thus tackling the SSB problem for modeling RFR prediction. 
Nevertheless, the $p_{cvr}$ is small practically, thus $p_{rfr}$ (i.e. $p_{cvrfr}$ divided by $p_{cvr}$) may be larger than 1, and then arise numerical instability. In our entire space modeling, the $p_{cvrfr}$ is computed as $p_{cvr}*p_{rfr}$, avoiding the problem of numerical instability. Furthermore,  the embedding layers are shared within the bottom layers,  which contribute most of the parameters in models, and learning it needs a huge number of samples. As such, the clicked samples (largely outnumber the conversion samples) are employed to help induce the RFR prediction model, alleviating the DS problem. Consisting of CVR and CVRFR prediction tasks, the learning objective is designed as follows:
\begin{equation}
    \mathcal{L} = - \sum_{\mathbf{x}}l( p(y=1|\mathbf{x}),y) 
    -\sum_{\mathbf{x}}l(p(z=1, y=1|\mathbf{x}),y\&z)
    \label{eq: esmm}
\end{equation}
where $y\&z$ returns 1 if $y=1$ and $z=1$, and 0 otherwise. Concretely, it corresponds to towers 2 and 3 and losses 2 and 3 in Fig.~\ref{fig: ECAD}. However, these are still CDF problems not yet been handled.
\vspace{-0.5em}
\subsubsection{Cascade Delayed Feedback Modeling}
The CDF problem in our task refers to that the delayed feedback exists both in the sequentially dependent CVR and RFR prediction tasks. As such, it needs a longer attribution window to observe the corresponding labels for simultaneously modeling the CVR and RFR predictions using entire space modeling than individually achieving these two tasks. Hence, the CDF problem significantly harms the data freshness for training the predictor to capture the rapid distribution shift of data in the recommender system. To tackle this problem, we carefully design auxiliary tasks for incorporating the conversion and refund time to advance the generalization performance of predictors in a multi-task learning framework. 

Inspired by Defer~\cite{gu2021real}, we employ the conversion and refund time contained within data for alleviating the delayed feedback problem in CVR and RFR prediction, respectively. Specifically, the $n+m$ towers are built to predict whether the conversion sample will convert within the predefined time windows $W_1, W_2, \cdots, W_n$ (e.g. $p(w<W_i|y=1,\mathbf{x}) (i: 1\to n)$), and whether the conversion but refunded sample will refund within the predefined time windows $V_1, V_2, \cdots, V_m$ (e.g. $p(v<V_i|z=1,y=1,\mathbf{x})(i: 1\to m)$). To train the model with all samples, these probabilities are further multiplied by $p(y=1|\mathbf{x})$ or $p(z=1, y=1|\mathbf{x})$ to obtain $p(w<W_i, y=1|\mathbf{x})$ and $p(v<V_i, z=1, y=1|\mathbf{x})$.
Denote the label that whether sample x converts/refunds within $W_i$/$V_i$ after click/conversion as $y_i$/$z_i$, Eq.~\eqref{eq: esmm} can be reformulated for achieving the CDF problem as follows:
\begin{equation}
\begin{aligned}
    \mathcal{L} =& -\sum_{\mathbf{x}}\sum_i l( p(w<W_i, y=1|\mathbf{x}),y_i) -\sum_{\mathbf{x}}l( p(y=1|\mathbf{x}),y)\\&
    -\sum_{\mathbf{x}}\sum_il( p(v<V_i, z=1, y=1|\mathbf{x}),y\&z_i) \\& - \sum_{\mathbf{x}}l( p(z=1, y=1|\mathbf{x}),y\&z) 
\end{aligned}
\label{eq: eacd-de}
\end{equation}
In concrete, this model corresponds to towers 1\textasciitilde4 and losses 1\textasciitilde4 in Fig.~\ref{fig: ECAD}. 
However, it cannot utilize the recently clicked samples whose CVR label $y$ has not been determined for learning the RFR prediction. To tackle this problem, we redesign the auxiliary tasks for the delayed RFR prediction model to excavate and leverage both the conversion and refund time for employing the recently clicked samples to boost the generalization performance. Specifically, we additionally add $n + n*m $ towers for the delayed RFR prediction model, in which $n$ heads predict whether the conversion but refunded sample will convert within the predefined time windows $W_1, W_2, \cdots, W_n$ (e.g. $p(w<W_i|z=1,y=1,\mathbf{x}) (i:1\to n$)) while other $n*m$ heads predict whether these samples will convert and refund with the predefined time windows $W_1, W_2, \cdots, W_n$ and $V_1, V_2, \cdots, V_m$ (i.e. $p(v<V_j,w<W_i|z=1,y=1,\mathbf{x})(i:1\to n,j:1\to m)$), respectively. Then these probabilities are multiplied by  $p(z=1, y=1|\mathbf{x})$ to model all the samples. Thus,  Eq.~\eqref{eq: eacd-de} is further schemed as follows:
\begin{equation}
\begin{aligned}
    \mathcal{L} =& -\sum_{\mathbf{x}}\sum_i l( p(w<W_i, y=1|\mathbf{x}),y_i) -\sum_{\mathbf{x}}l( p(y=1|\mathbf{x}),y)\\&
    -\sum_{\mathbf{x}}\sum_il(p(v<V_i, z=1, y=1|\mathbf{x}),y\&z_i) \\&
    -\sum_{\mathbf{x}}\sum_il(p(w<W_i, z=1, y=1|\mathbf{x}),y_i\&z) \\&
    -\sum_{\mathbf{x}}\sum_i\sum_jl( p(v<V_j, w<W_i, z=1, y=1|\mathbf{x}),y_i\&z_j) \\&
    - \sum_{\mathbf{x}}l( p(z=1, y=1|\mathbf{x}),y\&z) 
\end{aligned}
\label{eq: ECAD}
\end{equation}
However, the total number of towers in the entire model would increase from $n+m+2$ to $n*m +2n+m+2 $ when there are $n$ and $m$ pre-defined windows in the CVR and RFR tasks respectively,  corresponding to Eq.~\eqref{eq: eacd-de} and Eq.~\eqref{eq: ECAD}. The excessive number of towers results in the model being too complex, thus harming its practicality. To address this problem, our ECAD further cuts $n*m$ towers that predict $p(v<V_j,w<W_i|z=1,y=1,\mathbf{x})$ and  approximates these output  using $ p(v<V_i | z=1, y=1, \mathbf{x})$ and $ p(w<W_j, | z=1, y=1, \mathbf{x})$ as follows:
\begin{equation}
\begin{aligned}
   p (v<V_i, &w<W_j |  z=1, y=1, \mathbf{x}) = \\ & p(v<V_i | z=1, y=1, \mathbf{x})*p(w<W_j, | z=1, y=1, \mathbf{x})
\end{aligned}
\label{eq: ecad-lite}
\end{equation}
We hypothesize conversion and refund time are independent for conversion but refunded samples,  thus this equation holds and its effectiveness is demonstrated in later experiments.
In this way, the number of towers is reduced to $2n+m+2$, which largely improves applicability, and we name this simplified method as ECAD-Lite. Note that samples near the end of the training date may not have all $n+m+2$ labels (i.e. $y_1,y_2,\cdots,y_n,z_1,z_2,\cdots,z_m,y,z$), ECAD only updates corresponding parameters according to the observed labels.

\begin{table*}[!tb]
\vspace{-1em}
\caption{Performance of compared models on the production dataset in three tasks: CVR, RFR, and ECVR prediction. The best performance in each setting is bold-faced. $\circ/\bullet$  indicates that ECAD is statistically worse/better than the compared method by student pairwise $t$-test at 95\% confident level.}
\vspace{-1em}
\centering
\tabcolsep=0.9mm
\renewcommand\arraystretch{1}
\begin{tabular}{c|cccc|cccc|cccc}
    \hline 
    Tasks  & \multicolumn{4}{c|}{CVR Prediction}& \multicolumn{4}{c|}{RFR Prediction}& \multicolumn{4}{c}{ECVR Prediction}\\ \hline
 Metrics  &  \multicolumn{1}{c}{AUC} & \multicolumn{1}{c}{RI-AUC} & \multicolumn{1}{c}{PR-AUC} & \multicolumn{1}{c|}{RI-PR-AUC} &  \multicolumn{1}{c}{AUC} & \multicolumn{1}{c}{RI-AUC} & \multicolumn{1}{c}{PR-AUC} & \multicolumn{1}{c|}{RI-PR-AUC}&  \multicolumn{1}{c}{AUC} & \multicolumn{1}{c}{RI-AUC} & \multicolumn{1}{c}{PR-AUC} & \multicolumn{1}{c}{RI-PR-AUC}\\ \hline
   CVR-Base & 0.8568$\bullet$	& 0.00\% &	0.0573$\bullet$ &	0.00\% & \multicolumn{1}{c}- & \multicolumn{1}{c}-& \multicolumn{1}{c}- &-&-&-&-&-\\
   RFR-Base&-&-&-&- &0.6705$\bullet$	& 0.00\%	& 0.3309$\bullet$	& 0.00\% &-&-&-&-\\
   ECVR-Base -& -&- & -& -&-&-&-&- &  0.8582$\bullet$ &	0.00\% &	0.0458$\bullet$	& 0.00\%\\\hline
   IM &0.8568$\bullet$ &	0.00\% &	0.0573$\bullet$	& 0.00\%	 &0.6705$\bullet$ &	0.00\%	& 0.3309$\bullet$	& 0.00\%	 & 0.8593$\bullet$	& 21.57\%	& 0.0472$\bullet$ &	40.00\%\\
   IM-Defer & 0.8589\,\,\,	& 47.73\% &	\bf 0.0589\,\, &	\bf 57.14\%	 & 0.6716$\bullet$	& 7.80\%	& 0.3320$\bullet$ &	10.89\%	& 0.8596$\bullet$	& 27.45\% &	0.0469$\bullet$	& 31.43\%\\
   ESMM & 0.8565$\bullet$ &	-6.82\%	& 0.0568$\bullet$ &	-17.86\%	 & 0.6747$\bullet$	& 29.79\%	& 0.3326$\bullet$	& 16.83\%	& 0.8591$\bullet$ &	17.65\%	& 0.0466$\bullet$	& 22.86\%\\
   ECAD-De & 0.8584$\bullet$	& 36.36\%	& 0.0585\,\,\, & 	42.86\% &	0.6763$\bullet$	& 41.13\%	& 0.3359$\bullet$ &	49.50\%	& 0.8610$\bullet$	& 54.90\%   &	0.0473$\bullet$ & 	42.86\% \\
   ECAD-Lite &0.8587\,\,\, &	43.18\% & 0.0585\,\,\, &	42.86\%	& 0.6784$\bullet$	& 56.03\%	& 0.3358$\bullet$ &	48.51\% &	0.8616\,\,\,	 & 66.67\%	& 0.0475\,\,\, &	48.57\% \\
   ECAD & \bf 0.8595\,\,	& \bf 61.36\%	& 0.0587\,\,\,	& 50.00\% &	\bf 0.6800\,\,	&  \bf 67.49\%	& \bf 0.3369\,\, &	\bf  59.41\%	& \bf 0.8621\,\, &	\bf 76.47\% &	\bf 0.0480\,\,	&\bf 
 62.86\%\\\hline
   ESMM-Oracle & 0.8612\,\,\,	& 100.00\% &	0.0601\,\,\,	& 100.00\%	& 0.6846\,\,\, &	100.00\% &	0.3410\,\,\, &	100.00\% &	0.8633\,\,\,	& 100.00\% & 	0.0493\,\,\,	& 100.00\%\\\hline
\end{tabular}
\vspace{-1.5em}
\label{res: main}
\end{table*}

\section{Experiment}
In this section, we conduct extensive experiments on the offline dataset and online A/B testing to validate the superiority of ECAD.

\subsection{Experimental Setup}
\subsubsection{Dataset}
As there is no public dataset for ECVR prediction under the CDF problem,  we collect click traffic logs of 11 days from Alibaba’s recommender system to build the production dataset with 0.56 billion samples, each with 209 features (e.g user and item features). Furthermore, the number of conversions, refunds, and effective conversions, are 3.75, 0.61, and 3.14 million, respectively.
The samples in the first 10 days and 11th day are employed for training and testing respectively, and we randomly partition the testing set into 10 parts and report average evaluation results.
In addition, the attribution windows for conversion and refund are both set as 3 days. For example, if a sample is clicked (converted) on 1st day, its CVR (RFR) label can be determined at the end of 3rd day. As such, for a sample clicked on 1st day, its ECVR label can be obtained at the end of 5th day.
\subsubsection{Baseline}
To perform a comprehensive comparison, we implemented and compared the following methods:   
(i) \textbf{CVR-Base}: A base model trained using the samples clicked before 2nd day from the last, predicting the $p_{cvr}$.
(ii) \textbf{RFR-Base}: A base model trained using the samples  which convert before 2nd day from the last, predicting the $p_{rfr}$.
(iii) \textbf{ECVR-Base}: A base model trained using the samples clicked before the 4th day from the last, predicting the $p_{ecvr}$.
(iv) \textbf{IM}: Independent modeling (IM) employs the CVR-Base and RFR-Base to achieve the ECVR prediction using Eq~\eqref{eq: ecvr}.
(v) \textbf{IM-Defer}: It first utilizes Defer~\cite{gu2021real} to independently perform CVR and RFR modeling, and then estimates the $p_{ecvr}$ using Eq.~\eqref{eq: ecvr}.  
(vi) \textbf{ESMM}~\cite{ma2018entire}: This model fulfills the CVR and RFR modeling in an entire space modeling manner. It is first trained using Eq.~\eqref{eq: esmm} on the samples clicked before the 4th day from the last and then predicts $p_{ecvr}$ using Eq.~\eqref{eq: ecvr}.
(vii) \textbf{ECAD}: The proposed ECAD method first employs both the conversion and refund time to advance the model using Eq.~\eqref{eq: ECAD}, and then predicts $p_{ecvr}$ via Eq~\eqref{eq: ecvr}.
(viii) \textbf{ECAD-De}: A degenerate version of ECAD, which is trained using Eq.~\eqref{eq: eacd-de}.
(ix) \textbf{ECAD-Lite}: A lite version of ECAD, which cuts down towers using Eq.~\eqref{eq: ecad-lite}.
(x) \textbf{ESMM-Oracle}: An oracle version of ESMM, i.e. it can access CVR and RFR labels of all training samples, which provide the upper bound performance using the entire space modeling.

All methods are implemented with Python 2.7 and Tensorflow 1.12 and follow an Embedding\&MLP architecture. Specifically, CVR-Base, RFR-Base, and ECVR-Base have one tower above the bottom layers for predicting the respective probability, while others have multiple towers on top of the shared bottom layers to predict the corresponding probability. For a fair comparison, all compared methods use the following configurations: the optimizer AdagradDecay~\cite{duchi2011adaptive}, the learning rate 0.05, the bath size 256, the bottom layers and tower are configured as embedding layers and fully-connected layers with hidden size $\{512,256,128\}$ respectively, and the activation functions Leaky ReLU~\cite{maas2013rectifier}. Following~\cite{zhang2022towards}, each method is trained with one epoch.
Note that although there are several different works that focus on delayed feedback modeling~\cite{ktena2019addressing,yasui2020feedback,yang2021capturing,hou2021conversion,yang2022generalized, chen2022asymptotically,yasui2022learning}, we do not employ them as compared methods as we focus on how to deal with the CDF problems, but not on proposing an effective method to achieve delayed feedback in a single event.

\subsubsection{Evaluation Metric}
Following previous works~\cite{gu2021real,yang2021capturing}, we adopt two widely used metrics for evaluating offline experimental results. The first metric is the area under the ROC curve (AUC) which indicates the probability that a positive sample is ranked higher than a negative one (0.1\% improvement on AUC in industrial datasets is deemed as significant~\cite{zhou2018deep,sheng2023joint}). The second metric is the area under the precision-recall curve (PR-AUC), which is more sensitive than AUC in scenarios where negative samples  significantly outnumber positive ones. Besides, We also report the relative improvement of methods over the base model in each task on AUC and PR-AUC~\cite{gu2021real}. Taking the relative improvement of ECAD on AUC (denoted as RI-AUC) in CVR prediction as an example, the $\text{RI-AUC}_{\text{ECDA}}$ can be computed as follows:
\begin{equation}
    \text{RI-AUC}_{\text{ECDA}} = \frac{\text{AUC}_\text{ECDA}-\text{AUC}_\text{CVR-Base}}{\text{AUC}_\text{ESMM-Oracle}-\text{AUC}_\text{CVR-Base}}
\end{equation}
Obviously, the method's performance improves as the relative improvement approaches 100\%. 
\subsection{Experimental Result}
We evaluate all compared approaches on the production dataset in three tasks (i.e. CVR, RFR, and ECVR prediction), and report the experimental results in Table~\ref{res: main}. From this table, we have the following observations:\\
(i) The proposed ECAD almost always manifests superior performance compared to other methods across various tasks and metrics except for the PRAUC metric in CVR prediction where it falls slightly behind IM-Defer, validating the superiority of ECAD.\\
(ii) For CVR prediction, the big performance gap between delayed feedback modeling methods (ECAD, ECAD-Lite, ECAD-De, and IM-Defer) and approaches without delayed feedback modeling  (IM and ESMM) proves the necessity for tackling the delayed feedback problem, which significantly compromises the data freshness for training the predictor.\\
(iii) For RFR prediction, the entire space modeling strategies (ECAD, ECAD-Lite, ECAD-De, and ESMM) clearly outperform the independent modeling techniques (IM, IM-Defer) since the DS and SSB problem limit the supervised information and harm the model generalization, respectively. Besides, the results that IM-Defer is outperformed by ESMM demonstrate that the DS and SSB problems are more prominent for degrading the model performance than delayed feedback problems in RFR prediction. \\
(iv) For ECVR prediction, ECVR-Base loses to all other approaches, which confirms the effectiveness of our strategy for ECVR prediction by decomposing it into two sub-tasks, alleviating the CDF problem and making full use of different knowledge between non-conversion and conversion but refunded samples.\\ 
(v) ECAD and ECAD-Lite generally achieve better performance than ECAD-De especially in the RFR prediction, which confirms the effectiveness of our proposed techniques that employ both conversion and refund time contained within data for achieving the CDF problem in the RFR prediction, thus boosting the  performance of predictors.  Furthermore,  there is a slight decrease in performance for ECAD-Lite compared to ECAD, which proves the validity of our probability approximation manner in Eq.~\eqref{eq: ecad-lite} for simplifying the model architecture.

\subsection{Online A/B Testing}
The proposed ECAD approach for ECVR prediction is an offline training method, and we conduct online A/B testing to evaluate its effectiveness on our production systems that apply offline training.
To get a stable conclusion, we observe the online experiments for 7 days, which showed that ECAD achieved a 5.21\% improvement in ECVR compared to the ECVR-Base model, which can bring huge benefits to the e-commerce platform. Furthermore, the ECAD model has also been deployed on one of our recommender systems.


\section{Conclusion}
In this paper, we have investigated the practical but unexplored ECVR prediction task by partitioning it into two sub-tasks (i.e. CVR and RFR prediction) and pointed out a novel, thorny but untouched problem of cascade delayed feedback (CDF), where conversion and refund events both suffer the delayed feedback. To tackle these challenges, we propose an Entire space CAscade Delayed feedback modeling (ECAD) approach that adopts an entire space modeling framework to address the SSB and DS problems in learning RFR predictors. Furthermore, the auxiliary tasks are carefully designed to exploit both the conversion and refund time contained within the data to advance the performance of the method, alleviating the CDF. Experimental results on both the offline production dataset and online A/B testing validate the superiority of the proposed method. ECAD has been deployed in one of the recommender systems in Alibaba, bringing significant improvement of ECVR. 

\bibliographystyle{ACM-Reference-Format}
\bibliography{ECAD}


\end{document}